\documentclass[preprint,amssymb,amsmath,aip,apl,noshowpacs,letterpaper]{revtex4-1}
\usepackage{amsmath}%
\usepackage{bm}%
\usepackage{graphicx}
\usepackage{color}

\newcommand{\ignore}[1]{}

\begin{document}

\title{Iron-chalcogenide FeSe$_{0.5}$Te$_{0.5}$ coated superconducting
tapes for high field applications}

\author{Weidong Si}
\email{wds@bnl.gov}
\author{Juan Zhou}
\author{Qing Jie}
\author{Ivo Dimitrov}
\author{V. Solovyov}
\author{P. D. Johnson}
\affiliation{Department of Condensed Matter Physics and Materials
Science, Brookhaven National Laboratory, Upton, NY 11973}
\author{J. Jaroszynski}
\affiliation{National High Magnetic Field National Laboratory,
Florida State University, 1800 E. Paul Dirac Drive, Tallahassee,
FL 32310}
\author{V. Matias}
\author{C. Sheehan}
\affiliation{Superconductivity Technology Center, Los Alamos
National Laboratory, Los Alamos, NM 87545}
\author{Qiang Li}
\email{qiangli@bnl.gov} \affiliation{Department of Condensed
Matter Physics and Materials Science, Brookhaven National
Laboratory, Upton, NY 11973}

\date{\today}
\begin{abstract}
The high upper critical field characteristic of the recently
discovered iron-based superconducting chalcogenides opens the
possibility of developing a new type of non-oxide high-field
superconducting wires. In this work, we utilize a buffered metal
template on which we grow a textured FeSe$_{0.5}$Te$_{0.5}$ layer,
an approach developed originally for high temperature
superconducting coated conductors. These tapes carry high critical
current densities ($>$1$\times10^4$A/cm$^2$) at about 4.2\,K under
magnetic field as high as 25\,T, which are nearly isotropic to the
field direction. This demonstrates a very promising future for
iron chalcogenides for high field applications at liquid helium
temperatures. Flux pinning force analysis indicates a point defect
pinning mechanism, creating prospects for a straightforward
approach to conductor optimization.
\end{abstract}

\pacs{
74.78.-w, 
74.70.Ad  
74.62.-c, 
74.62.Bf, 
}

\maketitle

High field applications of superconductors have been dominated by
Nb$_3$Sn, a material which allows magnetic fields up to 20\,T to
be achieved at 4.2\,K.\cite{Wilson83} However, Nb$_3$Sn wires
typically require a post-winding heat-treatment, which is a
technically-challenging manufacturing step. Although high
temperature superconducting oxides (HTS) offer excellent superconducting
properties,\cite{Malozemoff08} their characteristically high
anisotropies and brittle textures, in addition to the high
manufacturing costs, have limited their applications. The
newly-discovered iron-based superconductors are semi-metallic
low-anisotropy materials with transition temperatures, $T_c$'s, up
to 55\,K.\cite{Ren08} The combination of extremely high upper
critical fields $H_{c2}$(0) ($\sim$\,100\,T),\cite{Hunte08}
moderate anisotropies of $H_{c2}^{ab}/H_{c2}^c$
(1-8),\cite{Jaroszynski08} and high irreversibility fields,
$H_{irr}$,\cite{Yamamoto09} makes this class of superconductors
appealing for high field applications. Recently, high critical
current densities, $J_c$'s, have been reported in Co-doped
BaFe$_2$As$_2$\cite{Lee10} and SmFeAs(O,F)\cite{Moll10} systems.

Chalcogenides, which are a sub-class of iron-based
superconductors, hold several practical advantages over the
pnictides. Although the $T_c$'s of chalcogenides are typically
below 20\,K, they exhibit lower anisotropies $\sim$\,2 with
$H_{c2}$(0)'s approaching 50\,T.\cite{Fang10,Braithwaite10,Si09}
They also have the simplest structure among the iron-based
superconductors and contain only two or three elements excluding
the toxic arsenic, which greatly simplifies their syntheses and
handling. Attempts to make polycrystalline wires by the
powder-in-tube method have yielded very low $J_c$ values so
far.\cite{Ozaki11} However, the coated conductor technology, which
has been developed for the second-generation (2D) HTS
wires,\cite{Larbalestier01} can be adapted for FeSe since the
in-plane lattice constants of YBCO and FeSe are very close. Here
we use textured metal template, made by ion beam assisted
deposition (IBAD), to grow $c$-axis oriented layers of
chalcogenide FeSe$_{0.5}$Te$_{0.5}$. We found that these
superconducting tapes have superior high field performance and
nearly isotropic $J_c$'s above 25\,T at about 4.2\,K.

\begin{figure}[b]
\centering
\includegraphics[width=6.5cm]{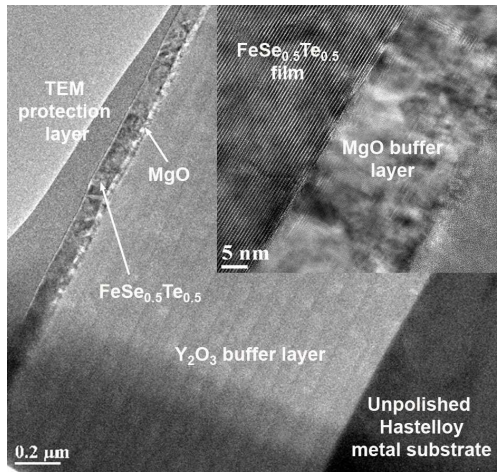}
\caption{Cross section HRTEM image of a FeSe$_{0.5}$Te$_{0.5}$
thin film on buffered metal template. The inset shows that
FeSe$_{0.5}$Te$_{0.5}$ inherits texture from the IBAD MgO layer.}
\label{fig_tem}
\end{figure}

We grow the FeSe$_{0.5}$Te$_{0.5}$ thin films by pulsed laser
deposition. The details of the growth conditions were published in
Ref. \onlinecite{Si09}. The films were deposited on single
crystalline LaAlO$_3$ (LAO) substrates and buffered metal
templates. The templates were manufactured in two steps. First, a
Y$_2$O$_3$ layer was made on unpolished Hastelloy by sequential
solution deposition to reduce the roughness of the tape surface,
then a bi-axially textured MgO layer was deposited on top by the
IBAD technique.\cite{Sheehan11} The very high tensile strength of
Hastelloy C-276 (0.8\,GPa) allows the composite conductor to
withstand the very high Lorentz force stresses produced by the
20-30\,T magnetic fields. Structural characterizations were
performed using high resolution transmission electron microscopy
(HRTEM). Resistivity and $J_c$ were measured by the standard
four-probe method.

Figure\,\ref{fig_tem} shows a cross-sectional HRTEM image of a
100\,nm FeSe$_{0.5}$Te$_{0.5}$ film on a buffered Hastelloy metal
substrate, consisting of a 1.3\,$\mu$m thick Y$_2$O$_3$
planarization layer and a bi-axially textured IBAD MgO layer
(including a 25\,nm homo-epitaxial MgO). The inset in
Fig.\,\ref{fig_tem} shows that the FeSe$_{0.5}$Te$_{0.5}$ film was
grown on the MgO layer with the $c$-axis perpendicular to the
substrate. X-ray diffraction experiments have also confirmed the
textured growth of FeSe$_{0.5}$Te$_{0.5}$, with in-plane and
out-of-plane textures about 4.5 and 3.5 degrees in full width half
maximum, respectively. However, the IBAD film has a lower zero
resistance $T_c^0$ ($\sim$\,11\,K) compared to the bulk
($\sim$\,14\,K), although the onset transition starts at
approximately the same temperature. The film on LAO has a
$T_c^0\sim$\,15\,K, about 1\,K above that of the bulk. \cite{Si09}
This may be because that MgO has a larger lattice mismatch with
FeSe$_{0.5}$Te$_{0.5}$ than LAO, which leads to more structural
defects.

\begin{figure}[t]
\centering
\includegraphics[width=7.5cm]{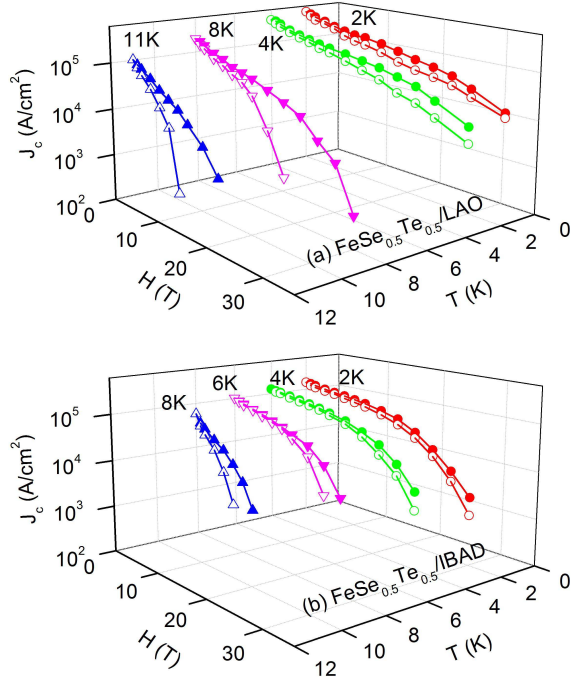}
\caption{$J_c$'s of FeSe$_{0.5}$Te$_{0.5}$ films on (a) LAO
substrate and (b) IBAD coated conductor at various temperatures
with magnetic field parallel (solid symbols) and perpendicular
(open symbols) to the $ab$ plane (tape surface).} \label{fig_Jc3D}
\end{figure}

Figure\,\ref{fig_Jc3D} shows the magnetic field dependence of
$J_c$ of films on both LAO and IBAD substrates at various
temperatures. The $J_c$ of films on LAO at $T\leq4$\,K is
$\sim5\times10^5$A/cm$^2$ in self-field, and remains above
$1\times10^4$A/cm$^2$ up to 35\,T, the maximum field we could
apply. Notably, the decrease of $J_c$ does not accelerate much at
high fields at liquid helium temperature, which is important for
high field applications. The $J_c$ decreases rather rapidly with
field at $T>8$\,K. Although the $J_c$'s of films on IBAD are lower
than those of films on LAO at the same temperature and field,
similar field behavior was observed. At $T\leq4$\,K, the
self-field $J_c$ is still as high as $2\times10^5$A/cm$^2$. In
comparison, the higher decreasing rates of $J_c$'s in the films on
IBAD were observed above 20\,T, but $J_c$'s still remain higher
than $\sim1\times10^4$A/cm$^2$ at 25\,T. Remarkably, in both
films, $J_c$'s are nearly isotropic with little dependence on
field direction at $T\leq4$\,K, a clear advantage for applications.

\begin{figure}[t] \centering
\includegraphics[width=6.5cm]{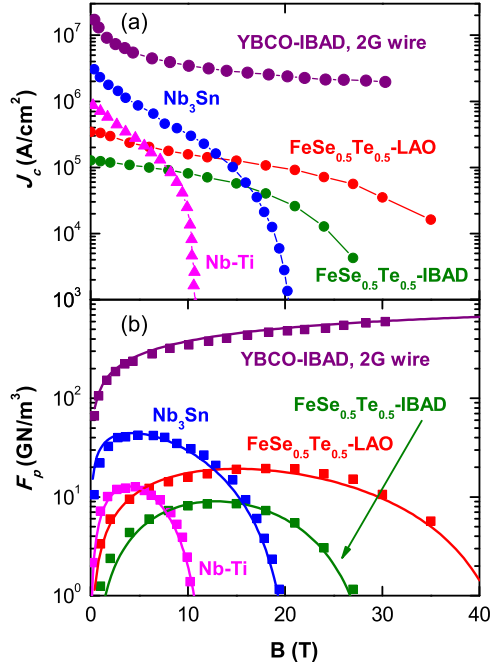}
\caption{(a) $J_c$ and (b) $F_p$ at about 4.2\,K of
FeSe$_{0.5}$Te$_{0.5}$ films compared with the literature data of
2G YBCO wire,\cite{Xu10} TCP Nb47Ti\cite{Cooley96} and
Nb$_3$Sn.\cite{Godeke75} For YBCO and FeSe$_{0.5}$Te$_{0.5}$ the
field direction is parallel to the $c$-axis. Solid lines in graph
(b) are Kramer's scaling approximations. } \label{fig_Comp}
\end{figure}

In Fig.\,\ref{fig_Comp} (a) and (b) we compare the field
dependence of $J_c$'s and volume pinning forces, $F_p=\mu_0H\times
J_c(B)$, for FeSe$_{0.5}$Te$_{0.5}$ films on LAO and IBAD
substrates with the literature data for 2G YBCO
wire,\cite{Xu10} thermo-mechanically processed Nb47Ti
alloy\cite{Cooley96} and small-grain Nb$_3$Sn wire\cite{Godeke75}
at about 4.2\,K. Clearly, FeSe$_{0.5}$Te$_{0.5}$ films exhibit
superior high field performance (above 20\,T) over those of low
temperature superconductors. HTS's currently present a great challenge for long-length wire production due to the rapid decrease of $J_c$ upon grain boundary misorientation, causing a subsequent increase in production costs.  That may not be as severe in FeSe$_{0.5}$Te$_{0.5}$. The IBAD substrates
have many low angle grain boundaries in the textured MgO template.
However, the IBAD FeSe$_{0.5}$Te$_{0.5}$ films are rather robust
with the self-field $J_c$ just a little lower than those of films
on LAO. It was reported that the grain boundary in a
Ba(Fe$_{1-x}$Co$_x$)$_2$As$_2$ system could reduce the $J_c$
significantly.\cite{Lee09} Our results seem to suggest that the
grain boundaries in iron chalcogenides may behave differently,
since they do not have a charge reservoir layer as in cuprates or
Ba(Fe$_{1-x}$Co$_x$)$_2$As$_2$, where carrier depletion occurs.
Measurements of FeSe$_{0.5}$Te$_{0.5}$ films grown on
bi-crystalline substrates are most desirable to provide direct
information on the misorientation angle dependence of $J_c$.

It is also possible that the relatively lower $J_c$'s in IBAD
films is simply due to the lower $T_c$'s compared to those of the
films on LAO, a result of the larger lattice mismatch between MgO
and FeSe$_{0.5}$Te$_{0.5}$. An additional buffer layer of CeO$_2$,
which has a better lattice match with FeSe$_{0.5}$Te$_{0.5}$, may
improve the $T_c$, and hence raise the $J_c$. On the other hand,
the elaborate oxide buffer structure, partially designed to
protect the metal template from oxidation for 2G HTS wires may not
be needed at all since FeSe$_{0.5}$Te$_{0.5}$ is made in vacuum.
Growing a FeSe$_{0.5}$Te$_{0.5}$ coating directly on textured
metal tapes may be possible, potentially simplifying the synthesis
procedure with a reduction of production costs.  Wire applications require much thicker (over several $\mu$m) films, which may be grown by using a more scalable deposition technique, such as a low-cost web-coating process for 2G HTS wire.

In Fig.\,\ref{fig_Comp} (b) we also show the Kramer's scaling law
approximation (solid line) $f_p\sim h^p(1-h)^q$ for different
types of superconductors at about 4.2\,K, where
$f_p=F_p/F_p^{max}$ is the normalized pinning force density and
$h=H/H_{irr}$ (${H}_{irr}$ is defined as the onset of zero resistance) is the reduced field.\cite{QL} We found that $q\sim2$ for all types of
superconductors, which is expected considering that the $(1-h)^2$ term
describes the reduction of the superconducting order parameter at
high field.\cite{Ekin10} The low field term $p\sim0.5$ ($h^{0.5}$)
was found for Nb$_3$Sn and YBCO and is associated with the
saturation regime, where $F_p^{max}$ changes little with the
pinning center density because flux motion occurs by shearing of
the vortex lattice, rather than by de-pinning.\cite{Good71} The
addition of BaZrO$_3$ nano-rods, which are very effective pinning
centers at 77\,K, resulted in a very minor pinning increase at
4.2\,K.\cite{Xu10} In contrast, the result of $p\sim1$ found in
the FeSe$_{0.5}$Te$_{0.5}$ system is similar to the one in Nb-Ti.\cite{Ekin10}
This is a strong evidence of point defect core pinning, most
likely from the inhomogeneous distribution of Se and
Te.  In the core pinning regime $F_p$ is a product of
the individual $F_p$ times the pinning center density. This means
that the $J_c$ of FeSe$_{0.5}$Te$_{0.5}$ can still be enhanced by
simply adding more defects to act as pinning centers. Due to the
short coherence length, we expect more pinning enhancement in
FeSe$_{0.5}$Te$_{0.5}$ before reaching the coupling limit.

In conclusion, we have grown $c$-axis oriented superconducting
FeSe$_{0.5}$Te$_{0.5}$ coated conductors by pulsed-laser
deposition. These tapes have a nearly isotropic $J_c$ above
10$^4$A/cm$^2$ under 25\,T of magnetic field at about 4.2\,K.
Pinning force analysis indicates a point defect pinning mechanism.
These properties show that FeSe$_{0.5}$Te$_{0.5}$ has a very
promising future for high field applications at liquid helium
temperatures.

This work was supported by the U.S. Department of Energy, Office
of Basic Energy Science, Materials Sciences and Engineering
Division, under Contract No. DE-AC02-98CH10886. IBAD templates
were provided by Los Alamos National Laboratory under funding from
the U.S. Department of Energy, Office of Electricity. A portion of
this work was performed at the National High Magnetic Field
Laboratory, which is supported by National Science Foundation
Cooperative Agreement No. DMR-0654118, the State of Florida, and
the U.S. Department of Energy.



\begin{thebibliography}{}
\bibitem{QL} 
${H}_{irr}$ is used as a scaling field, $h = H/{H}_{irr}$, rather than ${H}_{c2}$, the characteristic field in the original Kramer's scaling relation.  ${H}_{irr}$ has been found to give a better scaling behavior for $h$ in high-${T}_{c}$ cuprates in general.  In the case of iron chalcogenides, we found that scaling with either ${H}_{irr}$ or ${H}_{c2}$ gave similar exponents.
\end{thebibliography}
\end{document}